\documentclass[twocolumn,aps,amsmath,pre,floatfix]{revtex4}
\usepackage{graphicx} 
\usepackage{amssymb}
\usepackage{color}
\DeclareMathOperator{\Tr}{Tr}

\begin{document}
\title{Semi-flexible trimers on the square lattice in the full lattice limit}
\author{Pablo Serra}
\email{pablo.serra@unc.edu.ar}
\affiliation{Instituto de F{\'{\i}}sica Enrique  Gaviola,  CONICET
and Facultad  de  Matem{\'a}tica,
Astronom{\'{\i}}a, F{\'{\i}}sica  y  Computaci{\'o}n, Universidad  
Nacional  de  C{\'o}rdoba,  Argentina}
\author{W. G. Dantas}
\email{wgdantas@id.uff.br}
\affiliation{Departamento de Ci\^{e}ncias Exatas, EEIMVR, 
  Universidade Federal Fluminense, Volta Redonda, RJ,
  Brazil} 
  \author{J. F. Stilck}
  \email{jstilck@id.uff.br}
\affiliation{Instituto de F{\'{\i}}sica and National Institute of Science
  and Technology for Complex Systems, Universidade Federal Fluminense,
  Niter\'{o}i, RJ, Brazil}
\date{\today}

\begin{abstract}
    Trimers are chains formed by two lattice edges, and therefore three monomers. We consider trimers placed on the square lattice, the edges belonging to the same trimer are either colinear, forming a straight rod with unitary statistical weight, or perpendicular, a statistical weight $\omega$ being associated to these angular trimers. The thermodynamic properties of this model are studied in the full lattice limit, where all lattice sites are occupied by monomers belonging to trimers. In particular, we use transfer matrix techniques to estimate the entropy of the system as a function of $\omega$. The entropy $s(\omega)$ is a maximum at $\omega=1$ and our results are compared to earlier studies in the literature for straight trimers ($\omega=0$), angular trimers ($\omega \to \infty$) and for mixtures of equiprobable straight and angular trimers ($\omega=1$).
\end{abstract}

\maketitle

\section{Introduction  and definition of the model}
\label{intro}

The study of rod-like molecules has a long history in statistical mechanics.
Onsager \cite{o49} has argued that they should show orientional (nematic) order
at sufficiently high densities. Although his original argument was related to a
continuum system, later rods placed on lattices were also considered, such as in
the approximate study by Flory \cite{f56} somewhat later modified by Zwanzig
\cite{z63}.

So far, these lattice models of rods, where each rod ($k$-mer) is composed by
$k$ consecutive sites aligned in one of the directions of the lattice, were
much studied in the literature but the only case where an exact solution was 
found was for dimers ($k=2$) on two-dimensional lattices and in the full
lattice limit, when all lattice sites are occupied by rod monomers \cite{ktf61}. The full lattice dimer phase is isotropic.  On the square lattice, it was
found by Ghosh and Dhar \cite{gd07} that nematic ordering appears for rods
with $k \ge 7$ and at intermediate densities, at low and high densities 
the system is isotropic. The model has been studied extensively after the 
pioneering work by Ghosh and Dhar, mainly using computational simulations. With
increasing rod densities, the first transition from the isotropic to the nematic
phase, is continuous and its universality class for two-dimensional lattices 
was determined \cite{m08}. The second transition, where the system leaves 
the nematic phase and enters the high density isotropic phase is more difficult
to investigate through simulations: due to the high density of rods the moves
are very rare. An alternative simulational procedure allowed for more efficient
simulations in the high density region \cite{k13}, and recently there  were
shown arguments in favor of the possibility of this high density transition
being discontinuous on the square lattice \cite{s22}. Finally, the entropy 
of these system of rods in the limit of the full lattice, a generalization of 
the classical problem of the entropy of dimers for general size $k$ of the
rods, was also the subject of contributions in the literature. It was
estimated using transfer matrix techniques for trimers on the square lattice
by Ghosh et al \cite{g07}, computational simulations were used to produce
estimates, also on the square lattice for values of $k$ between 2 and 10 
\cite{p21} and transfer matrix techniques were applied also for 
$2 \le k \le 10$, leading to rather precise estimates for the full
lattice entropies \cite{r23}. An analytic approximation to this problem is 
the exact solution of the problem on the Husimi lattice \cite{rso22}, a
generalization of the solution of the problem on the Bethe lattice \cite{drs11}.
These solutions on the central region of treelike lattices may be seen as 
improved generalizations of simple mean field approximations.

In this paper we address a generalization of the athermal problem of trimers
on the square lattice mentioned above. If we allow the trimers to bend, they
now may be in two  configurations, and we may associate the energy 0 to the
extended configuration and the energy $\epsilon$ to the angular one, so that now
the model is thermal. The entropy per site of the system is now a function of
the  statistical weight of the bends $\omega=\exp[\epsilon/(k_BT)]$, where we
still are considering the full lattice limit. Besides the case $\omega=0$, which
corresponds to straight trimers and was discussed above, two other particular
cases of this problem were already studied before and discussed in the 
literature. In the limit $\omega \to \infty$ all trimers are in the angular 
configuration, and transfer matrix calculations on the square lattice were
used by Frob{\"o}se, Bonnemeier, and J{\"a}ckle to estimate the entropy 
\cite{f95}. When straight and angular $k$-mers are equiprobable ($\omega=1$),
similar techniques were used for $k$ in the range between 2 and 7 for the
full lattice limit \cite{d03}.

In the two limiting situations of the model which were already studied 
and described above ($\omega=0$ and $\omega \to \infty$), it is athermal, 
and here we consider the general case on the square lattice using transfer
matrix techniques to study the thermodynamic properties in the full lattice
limit. This is done defining the model on stripes of the square lattice of
finite widths $L$ in the horizontal $x$ direction, extending in the vertical
direction $y$. As discussed in more detail in \cite{r23}, the initial condition
at $y=0$ is fixed and the transverse boundary condition is periodic, so
that we are solving the model on cylinders with perimeters $L$. We then
proceed defining a transfer matrix for the model, and in the thermodynamic 
limit, where the height of the lattice diverges the properties of the model
are determined by the leading eigenvalue of this matrix. Among them, we 
calculate particularly the entropy per site $s(\omega)$. Finally, we extrapolate
the results for a finite sets of widths $L$ to the two-dimensional limit
$L \to \infty$ using finite size scaling techniques.

In the next section \ref{method} we describe the definition of the transfer 
matrix for the model and the method used to obtain the thermodynamic properties
of the system from it. In section \ref{results} we present the results and
compare them with previous ones were appropriate. Final comments and discussions
are presented in section \ref{final}

\section{Definition of the transfer matrix and calculation of the entropy}
\label{method}

\begin{figure}
    \centering
    \includegraphics{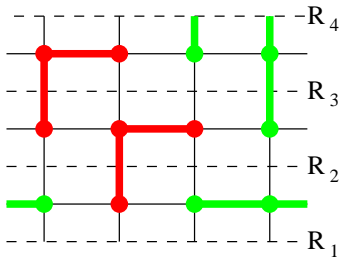}
    \caption{Section of a lattice with $L=4$.}
    \label{fig1}
\end{figure}

Consider a strip of transverse width $L$, with periodic boundary conditions 
in the transverse direction, such that every lattice site is occupied by a
monomer which belongs to a linear or bend trimer. The statistical weight of 
a linear trimer is unitary and the Boltzmann factor associated to a 
bend trimer is equal to $\omega$. A possible configuration of a section
of a strip with width $L=4$ is shown in figure \ref{fig1}.

The transfer matrix is defined in a similar way as in previous studies 
\cite{f95,g07,r23}. The states are defined by the configuration of the $L$
vertical edges of the lattice crossed by horizontal reference lines $R_i$ 
shown in the figure. Each time the transfer matrix is applied, $L$ sites are
added to de system, so that the line of the transfer matrix is defined by 
the configurations of the reference line $R_i$ and its column by the
configuration of the reference line $R_{i+1}$. The state may be represented by a
vector with  $L$ components. If the vertical edge which corresponds to a
component is not occupied by a trimer bond, the component is equal to zero,
otherwise, it will be equal to the number of monomers already included in the  
trimer (1 or 2). For example, the state associated to the reference line $R_1$ 
is $(0,0,0,0)$, and the state corresponding to $R_4$ will be $(0,0,1,2)$. The 
combinatorial problem of  building the transfer matrix is the following: 
given the state of reference line $R_i$, find all possible outputs for 
reference line $R_{i+1}$. To build the transfer matrix, we start with the 
state $(0,0,0,...0)$ and find all its possible outputs. Then we proceed finding
the second generation of output states. This iterative procedure ends when no
new states are generated. In general, a larger matrix could be defined and
the one we are considering is a particular block of this more general matrix, 
but as discussed in \cite{r23} there are evidences that the dominant eigenvalue
is contained in the block we are considering. 

\begin{figure}
    \centering
    \includegraphics[scale=0.7]{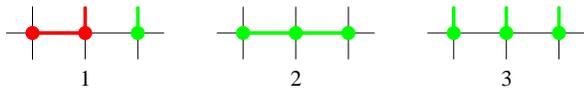}
    \caption{Output states of the state $(O,O,O)$ for $L=3$. }
    \label{fig2}
\end{figure}

For clearness, let us present in some detail the building of the transfer matrix
for $L=3$. We should notice that rotation and reflection symmetries are present
in the system, so we may reduce the size of the matrix using them. In figure
\ref{fig2} we see that the possible if we start with the state $(0,0,0)$:

1) we may proceed by placing an angular and starting a vertical trimer at the
three sites above, generating a state such as $(2,1,0)$, the statistical
weight associated is $\omega$, and there are six ways to do this;

2) a horizontal trimer may be placed on the three sites, the statistical weight
is unitary and the multiplicity is three. The output state is $(0,0,0)$;

3) three vertical trimers may be started at the sites, the statistical weight is
unitary, the multiplicity is one and the output state is $(1,1,1)$;

The only output state of $(1,1,1)$ is the state $(2,2,2)$ and this new state has
$(0,0,0)$ as its only output state, the multiplicity ans statistical weights are
unitary. We therefore, using the order of the four states as they appear in the
text above, have the following transfer matrix for the model on the $L=3$ strip:

\begin{equation}
T_3=
    \begin{pmatrix}
    3&6\omega&1&0\\
    \omega&1&0&0\\
    0&0&0&1\\
    1&0&0&0
    \end{pmatrix}
    .
\end{equation}

For periodic boundary conditions in the longitudinal direction of the strip, 
we find that the canonical partition function on a lattice of width $L$ 
and length $M$ may be written as:
\begin{equation}
Z=\Tr T_L^M=\sum_i \lambda_{i,L}^M,
\nonumber
\end{equation}
where the $\lambda_{i,L}$ are the eigenvalues of the transfer matrix $T_L$.
Therefore, in the thermodynamic limit $M \to \infty$, the free energy of
the model will be determined by the leading eigenvalue of the transfer matrix 
$\lambda_{1,L}$:
\begin{equation}
    f_L(T)=-\frac{k_BT}{L}\ln \lambda_{1,L}(\omega)
\end{equation}
and therefore the dimensionless entropy per site is given by
\begin{eqnarray}
    s_L(\omega)&=&\frac{-1}{k_B}\frac{\partial f_L}{\partial T} \nonumber \\
    &=&\frac{1}{L}\left(\ln \lambda_{1,L}+\omega \ln \omega 
    \frac{1}{\lambda_{1,L}}
    \frac{\partial \lambda_{1,L}}{\partial\omega}\right).
    \label{entrop}
\end{eqnarray}

In general, the transfer matrices are quite sparse and this favors using 
numerical procedures related to the power method to determine their leading
eigenvalue

As the width $L$ grows, the number of states $N_S$ increases and this
sets an upper limit to the widths we were able to consider, due to the
limitations of the computational resources at our disposal. Already considering
the symmetries, these numbers are displayed in table \ref{stm}. We  
considered widths up to $L=19$.

\begin{table}
\begin{center}
  \begin{tabular}{c c c c c c}
    \hline
    \hline
    $L$&$N_S$& & & $L$&$N_S$\\
    \hline
    3 & 4   &  & &12 & 7643  \\ 
    4 & 21  &  & &13 & 62415  \\
    5 & 39  &  & &14 & 173088 \\
    6 & 32  &  & &15 & 160544 \\
    7 & 198 &  & &16 & 1351983 \\
    8 & 498 &  & &17 & 3808083 \\
    9 & 409 &  & &18 & 3594014 \\
    10& 3210&  & &19 & 30615354\\
    11& 8418&  & &   &          \\  
    \hline
    \hline
  \end{tabular}
  \caption{Number of states (size of the transfer matrices) $N_S$ for
  the widths $L$ of the strips.}
  \label{stm}
\end{center}
\end{table}

\section{Results}
\label{results}

As described above, since we may calculate the largest eigenvalue of the 
transfer matrix as a function of $\omega$, we can obtain the entropy per
site $s(\omega)$ numerically using Eq. \ref{entrop}. As it happens for 
straight rods in the case $\omega=0$ \cite{r23},
the finite size behavior of those entropies can be separated in three sets of
values $\{s_L\}$, depending on the value of the remainder $R=L/k$, where $k=3$
is the number of monomers in the chain. This means that all entropy results
should be grouped in three sets of values, with $R=$ 0, 1, and 2. Each of those
sets tend to a thermodynamic limit, $s_{\infty}$, when $L\to\infty$, separately,
following finite size scaling behaviors of the form
\begin{equation}
    s_L^{(R)}(\omega)=s_{\infty}^{(R)}(\omega)+\frac{A(\omega)}{L^2}+o(L^{-2}),
    \label{relesc}
\end{equation}
also observed in the rigid chains limit and discussed in more detail in
\cite{r23}.

\begin{figure}[b]
\begin{center}
  \includegraphics[scale=0.65]{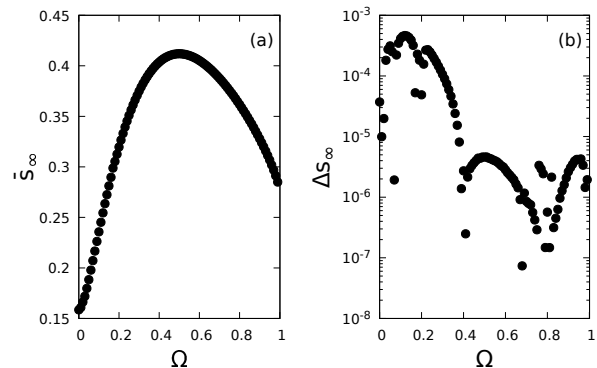}
  \caption{Extrapolated entropy values (a) and their estimated errors (b) as 
  functions of $\Omega$, calculated using Eqs. \ref{avg}.}
    \label{extraent}
\end{center}
\end{figure}

In the presentantion of the results, it is convenient to define the parameter
\begin{equation}
    \Omega=\frac{\omega}{1+\omega},
\end{equation}
whose range is between the values $\Omega=0$ (straight chains) and $\Omega=1$
(angular chains). We have used the BST 
method \cite{henkel} to extrapolate
the entropy values obtaining the estimates valid for the limit $L\to\infty$. The
BST was applied considering the relation \ref{relesc}, which implies to set the
value of the free parameter $w$ of this method to 2. Doing so, we 
determine the curves shown in figure \ref{extraent} considering the BST
extrapolation in each set of values determined by the remainder $R$. 
In that
figure we show the entropy as a function of the parameter $\Omega$.
The values obtained in each of those sets are quite close.
Actually, the differences among those values are within the uncertainty
furnished by the BST extrapolation procedure.

The circles shown in plot (a) of figure \ref{extraent} are the results of the
final extrapolation, obtained considering each of the sets statistically
independent and, for a given
value of $\Omega$, taking the values $s_i$ and its error, $\sigma_i$, evaluated
for each set, so that we get the final result for the extrapolation, 
$\overline{s}_{\infty}$ and its error, $\Delta S_{\infty}$ below:
\begin{eqnarray}
  \overline{s}_{\infty}&=&\frac{\sum_i s_i/\sigma_i^2}{\sum_i 1/\sigma_i^2}\nonumber\\
  \Delta s_{\infty}&=&\sqrt{\frac{1}{\sum_i 1/\sigma_i^2}}.
  \label{avg}
\end{eqnarray}
In plot (b) of figure \ref{extraent} the errors associated to the estimates 
of the entropies are depicted, we notice that, in general, they are smaller
in the region of higher values of $\Omega$, where angular trimers predominate.

In order to better evaluate this behavior, the table \ref{tavg} shows some of
the final extrapolated entropy values and their errors. Also, we highlight three 
special cases for this problem.
The case $\Omega=0$, previously mentioned as the limit where we have
only rigid trimers, was already studied by Gosh {\it{et. al}} \cite{g07} and
Rodrigues {\it{et. al}} \cite{r23}. In these works, that case was studied using
transfer matrices calculated for strips of widths up to $L=27$ and $L=36$,
respectively. Rodrigues {\it{et. al.}}, accomplish to reach such large widths
using a different method to determine the transfer matrix, called {\it Profile
Method}, which renders, in general, matrices with smaller dimensions. Still, our
result, obtained from the extrapolation of data considering widths up to $L=19$,
agrees with the more precise estimate reference \cite{r23}. Another particular
case mentioned before can be seen for the case $\Omega=1/2$, which means
$\omega=1$. This situation represents system of straight and angular trimers
with equal statistical weight in the full lattice limit. A previous result
obtained in \cite{d03} was calculated from transfer matrix approach considering
widths up to $L=12$ and is also consistent with the present value exhibited in
table \ref{tavg} for $\Omega=1/2$. Finally, our results do not include the case
$\Omega=1$, since the numerical calculation for the entropy from the general case
studied here displays strong fluctuations as long as we consider $\Omega\to 1$.
This case should be consider separately, as it was by Forb{\"o}se
{\it{et. al.}} \cite{f95}, whose result is shown in table \ref{tavg}.

\begin{table}[htbp!]
\begin{center}
  \begin{tabular}{ccccc}
    \hline
    \hline
    $\Omega$& & Extrapolated values& & Other results\\
    \hline
    0.0& &0.158539(37)  & & 0.15850494(19)\cite{r23}\\
    0.1& &0.23572(41)   & & \\
    0.2& &0.319592(49)  & & \\
    0.3& &0.37486(11)   & & \\
    0.4& &0.4036371(27) & & \\
    0.5& &0.4119467(46) & &  0.412010(20)\cite{d03}\\
    0.6& &0.4052574(28) & & \\
    0.7& &0.38767859(85)& & \\
    0.8& &0.36172496(57)& & \\
    0.9& &0.3278335(26) & & \\
    1.0& &              & &  0.276931500(50)\cite{f95}\\
    \hline
    \hline
  \end{tabular}
  \caption{Some values for the final extrapolated entropy as determined by Eq.
  \ref{avg}. Smaller errors are found for larger values of $\Omega$, where the 
  errors coming from the BST methods are minimum. The third column highlights 
  the values for the entropy calculated in previous studies considering special 
  cases, straight chains, $\Omega=0$, mixed trimers gas, $\Omega=1/2$ and 
  angular trimers, $\Omega=1$.}
    \label{tavg}
\end{center}
\end{table}

\begin{figure}[htbp!]
\begin{center}
  \includegraphics[scale=0.65]{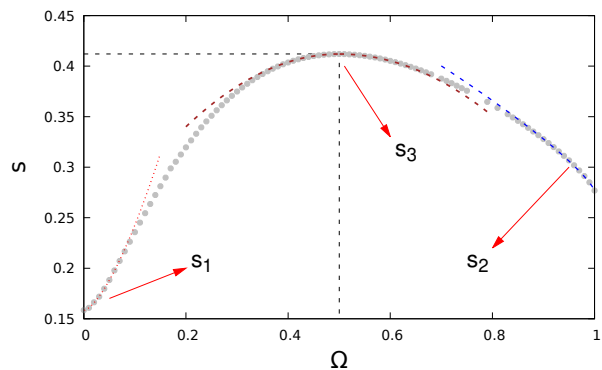}
  \caption{Same curve found in figure \ref{extraent}, but emphasizing three asymptotic regimes for the entropy behavior as a function of $\Omega$. The curves $s_1, s_2$ and $s_3$, explicited in Eqs.\ref{eq7},\ref{eq8} and Eq. \ref{eq9}. Each one of those regimes are related to the special cases mentioned in Table \ref{tavg}.}
    \label{asym}
\end{center}
\end{figure}

It is also interesting to find out how the curve shown in
figure \ref{extraent} behaves close to those three special points,
$\Omega=0, 1/2$ and $\Omega=1$. This is illustrated in figure \ref{asym}.
We have found three power-law kind of behaviors,
each one with a particular exponent. In the vicinity of $\Omega=0$, which means
to consider a system with only chains in the straight configuration, we may
see that entropy can be approximated as a function of $\Omega$ as
\begin{equation}
    s_1 \approx s_{\Omega=0}+A\Omega^{3/2},
    \label{eq7}
\end{equation}
where $s_{\Omega=0}$ is the value of the entropy for the rigid trimers case
and $A$ is a coefficient numerically estimated as $A \approx 2.67$. On the
other hand, close to the point $\Omega=1$ (angular trimers only) we found
that the entropy behavior tends to be like,
\begin{equation}
    s_2=s_{\Omega=1}+B(1-\Omega)^{4/5},
    \label{eq8}
\end{equation}
where $s_{\Omega=1}$ is the numerical value for the entropy when $\Omega=1$ and
$B\approx 0.32$. Finally, when we look at the function $s(\Omega)$ in the
neighborhood of $\Omega=1/2$, where we have straight and angular trimers
with equal weight, we get a quadratic relation for the entropy

\begin{equation}
     s_3=s_{\Omega=1/2}-C(1/2-\Omega)^2,
     \label{eq9}
\end{equation}
with $C \approx 0.7$. 

The fact that the maximum of the entropy is located exactly at $\Omega=1/2$,
which means $\omega=1$, is not really surprising, once this case favors no
particular arrangement (straight or angular), therefore corresponding to the
situation
where the number of configurations is maximum. Obviously, when we get away from
this point we are favoring one of the configurations. Since the case
of rigid trimers is the one with the least number of different configurations for
trimers occupying all sites of the lattice, this explains the asymmetry of the
curve around the point $\Omega=1/2$.

\section{Final discussion and conclusion}
\label{final}

The entropy of trimers fully occupying the square lattice was already estimated
before in three particular cases: when they are straight, composed by three
monomers on colinear sites \cite{g07,p21,r23}, when they are angular, so that 
the two edges which join the monomers are perpendicular \cite{f95} and for a 
mixture of equiprobable straight and angular chains \cite{d03}. Here we
generalize the  problem studying a canonical ensemble of trimers in straight
and angular configurations, associating an unitary statistical weight to the
first configuration and a weight $\omega$ to the second one. We thus, using
transfer matrix methods, obtain estimates for the entropy $s(\omega)$ for
$\omega$ in the range $[0,\infty]$, which are consistent with the earlier
results in the literature. 

The estimates are obtained from numerical diagonalization of the transfer
matrices of the model on strips of the square lattice with finite widths $L$ and
periodic boundary conditions in the transverse direction. The precision of the 
estimates is essentially determined by the largest width of the strips we are 
able to manage, since we use finite size scaling techniques to extrapolate
the results on strips of finite widths to the two-dimensional limit 
$L \to \infty$. As usual, the size of the transfer matrices grows exponentially 
with $L$ see table \ref{stm} and this leads to an upper limit $L=19$ we have
reached in this study. The computational part is divided in two stages: first 
the transfer matrix has to be built, and this is a combinatorial problem which
involves logical and integer variables only, and in the second stage the leading
eigenvalue of the transfer matrix, for a given value os the statistical weight
$\omega$, has to be determined. As the transfer matrix is very sparse, the 
determinations of the dominant eigenvalue is done using a variant of the 
power method. 

The extrapolation of the entropies of strips of finite widths $L$ to the 
two-dimensional limit $L \to \infty$ was done using the BST method 
\cite{henkel}, supposing that the asymptotic finite size scaling behavior
of the results is of the  form given by Eq. \ref{relesc}. This is discussed
in some detail for straight trimers in \cite{g07}. Although in many cases
the data are consistent with this behavior, as was already found for
straight $k$-mers for other values of $k$ \cite{r23}, it would still be
very interesting to have results for larger widths, in order to confirm this
asymptotic behavior. In particular, the amplitude $A$ in Eq. \ref{relesc}, 
as discussed in \cite{g07}, is related to the central charge of the phase, and
the present results do not allow precise estimates for it. 

Finally, we used an 
alternative procedure of extrapolation of the results for finite strips, 
adapted from the one employed by Frob{\"o}se et al. \cite{f95} in their study
of angular trimers. It relies the observation mentioned
above that the results for strips of widths $L$ should be divided in three 
groups, according to the value of the rest $R$ of the division of $L$ by 3. 
We notice, in general, that the results for the entropies  of the group $R=0$ 
approach the asymptotic value from above, while the one for $R=1,2$ 
approach it from below. We therefore consider the largest 3 or 6 widths $L$
and define the extrapolated entropy to be the mean value of the entropies
for these widths. The error will be given by the standard deviation of this 
set of entropies. For brevity, we will not present these results here,
but they were consistent with the ones obtained using the BST method. 

As is clear from the discussion above, it would be interesting to have
results for larger strips. One possibility we are considering is to apply
an alternative way to define the transfer matrix to this problem, which was
already used in \cite{r23} for straight $k$-mers with considerable success, in
the sense that it has lead, in that athermal problem, to smaller matrices, 
and therefore made it possible to solve that problem for larger widths than 
the ones accessible with the conventional procedure. In this grand canonical
transfer matrix procedure, at each application of the transfer matrix a variable
number of trimers is added to the system. We are presently working
on this problem.

\section{Acknowledgments}
It is a great pleasure for us to be part of this commemoration of Prof. Silvio
Salinas 80'th birthday. We thank him for the many opportunities in our
academic and personal lives where he has been not only the great teacher
and researcher we all know, but also a close friend. PS  acknowledges partial
financial support from CONICET and SECYT-UNC. and thanks the hospitality of
the Universidade Federal Fluminense, where  part of this manuscript was 
discussed and planned. This work used computational
resources from CCAD-UNC, which is part of 
SNCAD-MinCyT, Argentina.


\begin{thebibliography}{99}
\bibitem{o49}L. Onsager, Ann. N. Y. Acad. Sci. {\bf 51}, 627 (1949).
\bibitem{f56}P. J. Flory, Proc. R. Soc. {\bf 234}, 60 (1956)
\bibitem{z63}R. Zwanzig, J. Chem. Phys. {\bf 39}, 1714 (1963).
\bibitem{ktf61} P. Kasteleyn, Physica {\bf{27}}, 1209 (1961);P. Kasteleyn,
J. Math. Phys. {\bf{4}}, 287 (1963);
H. N. V. Temperley and M. E. Fisher, Philos. Mag. {\bf{6}}, 1061(1961); 
M. E. Fisher, Phys. Rev. {\bf{124}}, 1664 (1961).
\bibitem{gd07}A. Ghosh and D. Dhar, Europhys. Lett. {\bf 78}, 20003 (2007).
\bibitem{m08}D. A. Matoz-Fernandez, D. H. Linares, and A. J. Ramirez-Pastor,
Europhys. Lett. {\bf82}, 50007 (2008); D. A. Matoz-Fernandez, D. H. Linares,
and A. J. Ramirez-Pastor, J. Chem. Phys. {\bf 128}, 214902 (2008);  T. Fischer
and R. L. C. Vink,  Europhys.  Lett. {\bf 85}, 56003(2009); D. Matoz-Fernandez,
D. Linares, and A. Ramirez-Pastor, Physica A {\bf 387}, 6513  (2008).
\bibitem{k13}J. Kundu, R. Rajesh, D. Dhar, and J. F. Stilck, Phys. Rev. E
{\bf 87}, 032103 (2013).
\bibitem{s22}A. Shah, D. Dhar, and R. Rajesh, Phys. Rev. E {\bf 105},
034103 (2022).
\bibitem{g07} A. Ghosh, D. Dhar and J.L. Jacobsen, Phys. Rev. E
{\bf{75}}, 011115  (2007).
\bibitem{p21}P.M. Pasinetti, A.J. Ramirez-Pastor, E.E. Vogel and
G. Saravia, Phys. Rev. E {\bf{104}}, 054136 (2021). 
\bibitem{r23}L. R. Rodrigues , J. F. Stilck , and W. G. Dantas,
Phys. Rev. E {\bf 107}, 014115 (2023).
\bibitem{rso22} N. T. Rodrigues, J. F. Stilck, and T. J. Oliveira, Phys.
  Rev. E {\bf 105}, 024132 (2022).
\bibitem{drs11} D. Dhar, R. Rajesh, and J. F. Stilck, Phys. Rev. E {bf 84},
  011140(2011).
\bibitem{f95}K. Frob{\"o}se, F. Bonnemeier, and J. J{\"a}ckle, 
J. Phys. A: Math. Gen. {\bf 29} 485 (1996).
\bibitem{d03}W.G. Dantas and J.F. Stilck, Phys. Rev. E, {\bf 67}, 031803 (2003).
\bibitem{henkel} M. Henkel and G. Schutz, J. Phys. A {\bf{21}}, 2617 (1988).
\end{thebibliography}
\end{document}